\documentclass[12pt,a4paper]{article}
\usepackage{amsfonts,latexsym}
\usepackage{graphicx,color}
\usepackage{dcolumn}
\usepackage{graphicx}
\usepackage{amsmath}
\usepackage{amsfonts}
\usepackage{amssymb}

\renewcommand{\thefootnote}{\fnsymbol{footnote}}

\begin{document}

\vspace{12mm}

\begin{center}
{{{\Large {\bf Instability of Reissner-Nordstr\"{o}m black hole\\ in Einstein-Maxwell-scalar theory }}}}\\[10mm]

{Yun Soo Myung$^a$\footnote{e-mail address: ysmyung@inje.ac.kr} and De-Cheng Zou$^{a,b}$\footnote{e-mail address: dczou@yzu.edu.cn}}\\[8mm]

{${}^a$Institute of Basic Sciences and Department  of Computer Simulation, Inje University Gimhae 50834, Korea\\[0pt] }

{${}^b$Center for Gravitation and Cosmology and College of Physical Science and Technology, Yangzhou University, Yangzhou 225009, China\\[0pt]}
\end{center}
\vspace{2mm}

\begin{abstract}
The scalarization of Reissner-Nordstr\"{o}m  black holes  was  recently proposed in the Einstein-Maxwell-scalar  theory.
Here, we show that the appearance of the scalarized Reissner-Nordstr\"{o}m black hole is closely related to the Gregory-Laflamme
instability of the Reissner-Nordstr\"{o}m black hole without scalar hair.
\end{abstract}
\vspace{5mm}

\vspace{1.5cm}

\hspace{11.5cm}
\newpage
\renewcommand{\thefootnote}{\arabic{footnote}}
\setcounter{footnote}{0}


\section{Introduction}
Recently, the scalarized black hole solutions were found from Einstein-scalar-Gauss-Bonnet (ESGB)  theories~\cite{Doneva:2017bvd,Silva:2017uqg}.
We note that these black holes with scalar hair  are connected  to the appearance of  instability for the Schwarzschild black hole without scalar hair.
Interestingly, the instability of Schwarzschild black hole in ESGB theory is regarded  as not the tachyonic instability
but the Gregory-Laflamme (GL) instability~\cite{Gregory:1993vy} by comparing it with the instability of the Schwarzschild black hole in the Einstein-Weyl gravity~\cite{Myung:2018iyq}.

The notion of the GL instability comes from the three observations~\cite{Whitt:1985ki,Myung:2013doa,Lu:2017kzi,Stelle:2017bdu}: i) The instability is based on the $s(l=0)$-mode perturbations for scalar and tensor fields.
ii) The linearized equation includes an effective mass term, providing that the potential develops negative region near the black hole horizon but it becomes positive after crossing the $r$-axis.
iii) The instability of a black hole without hair  is closely related to the appearance of a newly black hole with hair.

More recently, a scalarization of the Reissner-Nordstr\"{o}m (RN) black hole  was proposed in the Einstein-Maxwell-scalar (EMS) theory which is considered  as a simpler theory than the ESGB theory~\cite{Herdeiro:2018wub}.
We note that the scalarized black holes were found in the Einstein-scalar-Born-Infeld theory~\cite{Stefanov:2007eq,Doneva:2010ke}, regarded  as a generalized EMS theory.
The EMS theory includes three  physically propagating modes of scalar, vector, and tensor.
In this case, the instability of RN black hole is determined solely by the linearized scalar equation because the RN black hole is stable against tensor-vector perturbations, as found in the Einstein-Maxwell theory~\cite{Zerilli:1974ai,Moncrief:1974gw,Moncrief:1974ng,Moncrief:1975sb}.

In this work, we wish to show that the appearance of the scalarized RN black hole is closely associated with the GL
instability of the RN black hole without scalar hair.  Here, the GL instability will be determined by
solving the linearized scalar equation. This will indicate an important connection between scalarized  RN  black holes and GL instability of RN black holes.

The organization of our work is as follows. We introduce the EMS theory  and its linearized theory around the RN black hole background in section 2.
In section 3, we perform the stability analysis for the RN black hole based on  the linearized scalar equation (\ref{pertur-eq}).
Mainly, we derive the GL instability bound (\ref{inst-b}). We solve the static linearized  equation (\ref{phi-r}) to confirm the threshold of the instability $\alpha_{\rm th}$ as well as to obtain $n=0,1,2\cdots$ scalarized RN black holes in section 4. In section 5, we explore what the GL instability is. Section 6 is devoted to obtaining  a scalarized RN black hole by solving  the four equations (\ref{neom1})-(\ref{neom4}) numerically. It indicates that the appearance of the scalarized RN black hole is closely related to the GL
instability of the RN black hole without scalar hair. Also, we obtain scalarized RN black holes for the quadratic  coupling of $\alpha \phi^2 $ for comparison to the exponential coupling $ e^{\alpha \phi^2}$.  Finally, we will describe our main results in section 7.

\section{EMS and its linearized theory} \label{sec1}

The EMS theory  is given by~\cite{Herdeiro:2018wub}
\begin{equation}
S_{\rm EMS}=\frac{1}{16 \pi}\int d^4 x\sqrt{-g}\Big[ R-2\partial_\mu \phi \partial^\mu \phi-V_\phi-e^{\alpha \phi^2} F^2\Big],\label{Action1}
\end{equation}
where $\phi$ is the scalar field  with a potential $V_\phi$, $\alpha$ is a positive  coupling constant, and $F^2=F_{\mu\nu}F^{\mu\nu}$ is the Maxwell kinetic term.
Here we choose $V_\phi=0$ for simplicity. This theory implies that three of scalar, vector, and tensor are physically dynamical fields. It is noted that  a different dilaton coupling of $e^{-2\alpha_0 \phi}$ was introduced  for the Einstein-Maxwell-dilaton theory originating from a low-energy limit of string theory~\cite{Hirschmann:2017psw,Pacilio:2018gom}.  Moreover, a quadratic  coupling of $\alpha \phi^2$ will be considered as the other model to reveal scalarized charged black holes in section 7.

Now, let us derive  the Einstein  equation from the action (\ref{Action1})
\begin{eqnarray}
 G_{\mu\nu}=2\partial _\mu \phi\partial _\nu \phi -(\partial \phi)^2g_{\mu\nu}+2e^{\alpha \phi^2}T_{\mu\nu}, \label{equa1}
\end{eqnarray}
where $G_{\mu\nu}=R_{\mu\nu}-(R/2)g_{\mu\nu}$ is  the Einstein tensor and  $T_{\mu\nu}=F_{\mu\rho}F_{\nu}~^\rho-F^2g_{\mu\nu}/4$   is the Maxwell energy-momentum tensor.
The Maxwell equation is given by
\begin{equation} \label{M-eq}
\nabla^\mu F_{\mu\nu}-2\alpha \phi\nabla^{\mu} (\phi)F_{\mu\nu}=0.
\end{equation}
 The scalar equation takes the form
\begin{equation}
\square \phi -\frac{\alpha}{2} e^{\alpha \phi^2}F^2 \phi=0 \label{s-equa}.
\end{equation}

Considering   $\bar{\phi}=0$ and electrically charged  $\bar{A}_t=Q/r$, one finds the RN  solution from (\ref{equa1}) and (\ref{M-eq})
\begin{equation} \label{ansatz}
ds^2_{\rm RN}= \bar{g}_{\mu\nu}dx^\mu dx^\nu=-f(r)dt^2+\frac{dr^2}{f(r)}+r^2d\Omega^2_2
\end{equation}
with the metric function
\begin{equation}
f(r)=1-\frac{2M}{r}+\frac{Q^2}{r^2}.
\end{equation}
Here, the outer  horizon is located at $r=r_+=M+\sqrt{M^2-Q^2}=M(1+\sqrt{1-q^2})$ with $q=Q/M$, while the inner horizon is at $r=r_-=M(1-\sqrt{1-q^2})$.
It is worth noting  that (\ref{ansatz})  dictates   a charged black hole solution without scalar hair.
 We stress  that the RN solution (\ref{ansatz}) is a black hole solution to the EMS theory for any value of $\alpha$.
Hereafter we are interested in the outer horizon.

In order to explore the stability analysis, one has to obtain  the linearized theory which describes the metric perturbation   $h_{\mu\nu}$, vector perturbation $a_\mu$ and scalar perturbation $\varphi$ propagating around the RN background (\ref{ansatz}) denoting by  $\bar{{}}$ (overbar).  By linearizing  (\ref{equa1}), (\ref{M-eq}), and (\ref{s-equa}), we find three linearized equations as
\begin{eqnarray}
&&  \delta G_{\mu\nu}(h) = 2\delta T_{\mu\nu}, \label {l-eq1}\\
&&  \bar{\nabla}^\mu f_{\mu\nu}=0, \label{l-eq2} \\
&& \Big(\bar{\square}+ \alpha \frac{Q^2}{r^4}\Big) \varphi= 0, \label{l-eq3}
\end{eqnarray}
where the linearized Einstein tensor $\delta G_{\mu\nu}$, the linearized energy-momentum tensor $\delta T_{\mu\nu}$, and the linearized Maxwell tensor $f_{\mu\nu}$ are given by
\begin{eqnarray}
  \delta G_{\mu\nu} &=& \delta R_{\mu\nu}-\frac{1}{2} \bar{g}_{\mu\nu} \delta R -\frac{1}{2} \bar{R} h_{\mu\nu}, \label{l-G} \\
  \delta T_{\mu\nu} &=&\bar{F}_{\nu}~^\rho f_{\mu\rho}+\bar{F}_{\mu}~^\rho f_{\nu\rho}-\bar{F}_{\mu\rho}\bar{F}_{\nu\sigma}h^{\rho\sigma}\nonumber   \\
                    &+& \frac{1}{2}(\bar{F}_{\kappa\eta}f^{\kappa\eta}-\bar{F}_{\kappa \eta}\bar{F}^\kappa~_\sigma h^{\eta\sigma})\bar{g}_{\mu\nu}-\frac{1}{4}\bar{F}^2h_{\mu\nu},
                    \label{l-T}\\
  f_{\mu\nu} &=& \partial_\mu a_\nu-\partial_\nu a_\mu. \label{l-F}
\end{eqnarray}
We note that an effective mass term of $-\alpha Q^2/r^4$ in (\ref{l-eq3}) is replaced by $-2\lambda^2M^2/r^6$ in the ESGB theory~\cite{Myung:2018iyq}.
Here the scalar coupling constant `$\alpha>0$' plays the role of a mass-like parameter.

\section{Instability of  RN black hole }

In analyzing  the stability  of the RN black hole in the EMS theory,
we first consider  the two linearized   equations (\ref{l-eq1}) and (\ref{l-eq2}) because two perturbations of metric $h_{\mu\nu}$ and vector $a_{\mu}$   are coupled.
Exactly, these correspond to the linearized equations for the Einstein-Maxwell theory.
For  the odd-parity perturbations, one found the Zerilli-Moncrief equation which describes two physical DOF propagating around the RN  background
~\cite{Zerilli:1974ai,Moncrief:1974gw}.
Also, the even-parity perturbations with two physical degrees of freedom (DOF) were studies in~\cite{Moncrief:1974ng,Moncrief:1975sb}.
It turns out that the RN black hole is  stable against these perturbations. In
this case,  a massless spin-2 mode starts with  $l=2$, while a massless spin-1 mode begins with $l=1$. The EMS theory provides 5(=2+2+1) DOF propagating around the RN background.

Now, we focus on  the linearized scalar equation (\ref{l-eq3}) which determines the stability of the RN black hole   in the EMS theory.
Introducing
\begin{equation} \label{scalar-sp}
\varphi(t,r,\theta,\chi)=\frac{u(r)}{r}e^{-i\omega t}Y_{lm}(\theta,\chi),
\end{equation}
and considering  a tortoise coordinate $r_*$ defined by $dr_*=dr/f(r)$, a radial equation of (\ref{l-eq3}) leads to the Schr\"{o}dinger-type equation
\begin{equation} \label{sch-2}
\frac{d^2u}{dr_*^2}+\Big[\omega^2-V(r)\Big]u(r)=0,
\end{equation}
where the scalar potential $V(r)$ is given by
\begin{equation} \label{pot-c}
V(r)=f(r)\Big[\frac{2M}{r^3}+\frac{l(l+1)}{r^2}-\frac{2Q^2}{r^4}-\alpha\frac{Q^2}{r^4}\Big].
\end{equation}
\begin{figure*}[t!]
   \centering
  \includegraphics{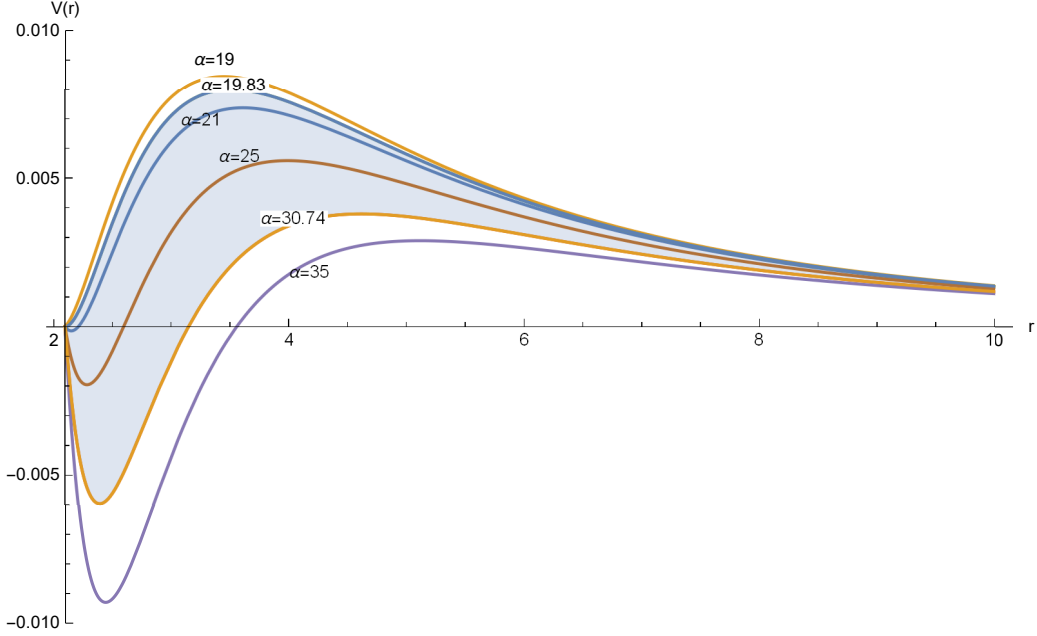}
\caption{The $\alpha$-dependent potentials as function of $r\in [r_+,\infty)$  for the outer horizon radius $r_+=2.09(q=Q/M=0.418)$ and $l=0$.  From the top, each curve represents the potential  $V(r)$  of a scalar field for the  parameter $\alpha=19$ (stable), 19.83 (positive definite potential: sufficient condition for stability), 21, 25, 30.74 (sufficient condition for instability), and 35 (unstable case), respectively. The potentials   have negative regions near the  horizon for $\alpha>19.83$. One conjectures that  the threshold of GL instability  occurs for  $\alpha_{\rm th}>19.83$.  }
\end{figure*}
In Fig. 1, we find the $\alpha$-dependent potentials for given $l=0$, $M=1.1$ and $q=M/Q=0.418$ (a non-extremal RN black hole).
The $s(l=0)$-mode is allowed for the scalar perturbation and it is regarded as an important mode to test the stability of the RN black hole.
Hereafter,   we consider this mode only.

A sufficient condition of $\int^\infty_{r_+} dr V(r)/f(r)<0$ for instability~\cite{BS,Dotti:2004sh}  leads to the bound as
\begin{equation}
\alpha> \alpha_{\rm in}(q)\equiv \frac{3}{q^2}-\frac{2q^2-3\sqrt{1-q^2}}{q^2}. \label{sc-in}
\end{equation}
 The first term  $3/q^2$  was found in analyzing the black hole dynamics in Einstein-Maxwell-dilaton theory~\cite{Hirschmann:2017psw}.

On the other hand, by observing the potential (\ref{pot-c}) carefully,  the positive definite potential without negative region  could be implemented  by imposing the bound
\begin{equation}
\alpha \le \alpha_{\rm po}(q)\equiv \frac{2(1-q^2)}{q^2}+\frac{2\sqrt{1-q^2}}{q^2},\label{c-po}
\end{equation}
which guarantees a stable RN black hole. This is called the sufficient condition for stability.
We note that  (\ref{sc-in}) is not a necessary and sufficient condition for the  instability.
Observing  Fig.~1 together with $q=0.418$, one finds that   two potentials with $\alpha=21,25$ between $\alpha_{\rm po}=19.83 $ and {\it $\alpha_{\rm in}=30.74$} develop  negative regions near the horizon, but they become positive after crossing the $r$-axis.

At this stage, we would like to mention that  such potentials exist around neutral black holes (black holes without charge) in higher dimensions  and the $S$-deformation has been used to confirm the stability of neutral black holes in higher dimensions~\cite{Kodama:2003kk}. We conjecture  that the GL instability may  occur for  $\alpha_{\rm th}> 19.83$, but the threshold of instability $\alpha_{\rm th}$ should  be determined explicitly  by  the numerical computations.
We expect that $\alpha_{\rm th}$  is located at the shaded region between $\alpha_{\rm in}$ and $\alpha_{\rm po}$.
Usually, if the potential $V$ derived from physically propagating modes is negative in some region, a growing perturbation can appear in the spectrum. This might indicate  an instability of the  black hole system under such perturbations.
However, this is not always true. Some potentials with negative region near the horizon  do not imply the instability. The criterion to determine whether a black hole
is stable or not  against the perturbation is whether the time-evolution of the perturbation is decaying or not.
The perturbed equation around a RN black hole can usually be described by the Schr\"{o}dinger-type equation (\ref{pertur-eq}),  where a growing mode like $e^{\Omega t}$ of the perturbation indicates the instability of the black hole. The absence of any  unstable  physical fields provides  a precise way of determining the stability of the black hole.

It suggests that  the RN black hole would be unstable for $\alpha>\alpha_{\rm th}=29.47$ with $q=0.418$,  while it is stable  for $19.83<\alpha<29.47$ showing  negative region near the horizon.  In the latter case,  the $S$-deformation method could provide a complementary result to  support  the stability of such black holes by finding the deformed potential~\cite{Kimura:2017uor,Kimura:2018eiv}.
\begin{figure*}[t!]
   \centering
   \includegraphics{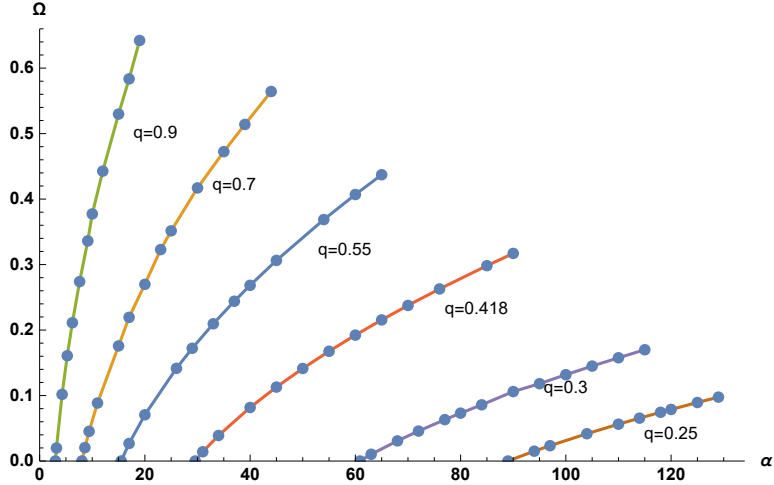}
\caption{Plots of unstable scalar modes ($\bullet$) on six different  curves with $q=\{0.25,0.3,0.418,0.55,0.7,0.9\}$.
The $y(x)$-axis denote $\Omega$ in $e^{\Omega t}$ (mass-like parameter $\alpha$). Here we observe that the thresholds $(\Omega=0$) of instability are located at $\alpha_{\rm th}(q)\approx \{88.98,60.69,29.47,15.46,8.019,2.995\}$.}
\end{figure*}
\begin{figure*}[t!]
   \centering
  \includegraphics{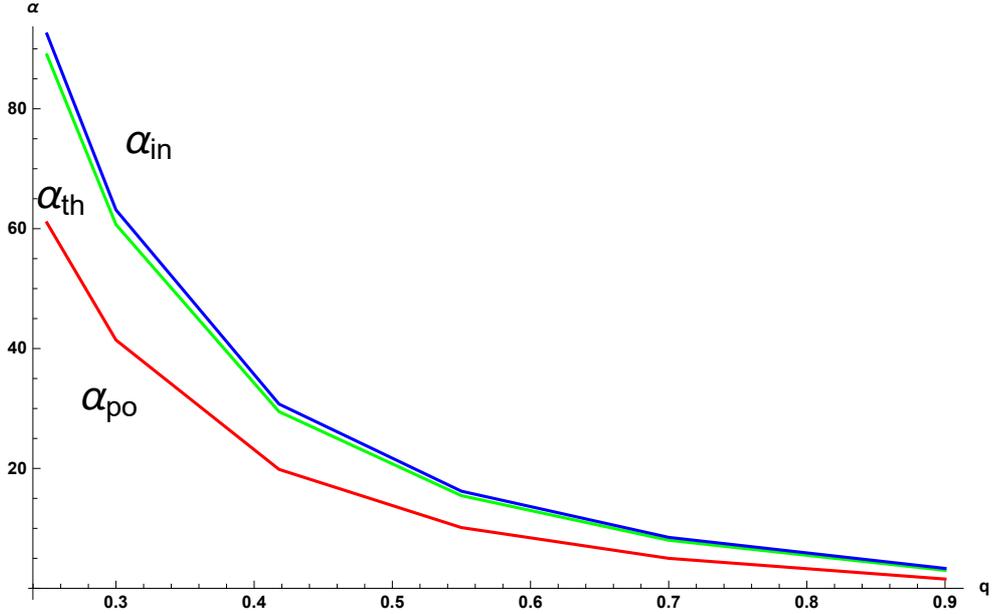}
\caption{Three $\alpha$-curves  as function of $q$. The upper blue curve represents $\alpha_{\rm in}(q)$ in (\ref{sc-in}) and the middle green curve indicates $\alpha_{\rm th}(q)$, while the lower red one denotes $\alpha_{\rm po}(q)$ (\ref{c-po}).
We find an inequality of $\alpha_{\rm po}(q)<\alpha_{\rm th}(q) <\alpha_{\rm in}(q)$. }
\end{figure*}

To determine the threshold of instability explicitly, one has to solve the second-order differential equation numerically
\begin{equation}\label{pertur-eq}
\frac{d^2u}{dr_*^2}-\Big[\Omega^2+V(r)\Big]u(r)=0,
\end{equation}
which  allows an exponentially growing mode of  $e^{\Omega t}(\omega=i\Omega) $ as  an unstable mode.
Here we choose two boundary conditions: a normalizable
solution of $u(\infty)\sim e^{-\Omega r_*}$  at infinity  and
a solution of $u(r_+)\sim \left(r-r_+\right)^{\Omega r_+}$  near the outer horizon.
Observing  Fig.~2,  we read off the threshold of instability $\alpha_{\rm th}(q)=\{88.98,60.69,29.47,15.46,8.019,2.995\}$
at $q=\{0.25,0.3,0.418,0.55,0.7,0.9\}\equiv\{\cdots\}$.
From Fig. 3,   one confirms that   the threshold of instability is located between  the sufficient condition for instability $\alpha_{\rm in}(q)\approx \{92.48,63.13,30.74,16.2,8.45,3.318\}$ at $ q=\{\cdots\}$
and the sufficient condition for stability $\alpha_{\rm po}(q)\approx \{68.98,41.42, 19.83, 10.13, 4.997, 1.545\}$ at  $q=\{\cdots\}$. This implies  an inequality  as
\begin{equation} \label{ineq-r}
 \alpha_{\rm po}(q) <\alpha_{\rm th}(q) <\alpha_{\rm in}(q),
\end{equation}
where $\alpha_{\rm po}(q)$ and $\alpha_{\rm in}(q)$ are given by (\ref{c-po}) and (\ref{sc-in}), while $\alpha_{\rm th}(q)$ is determined by solving (\ref{pertur-eq}) numerically.

Consequently, the GL instability bound for the RN black hole is given by
\begin{equation}
\alpha >\alpha_{\rm th}(q) \label{inst-b}
\end{equation}
which is considered as  one of our main results.
However, we could not determine  an explicit form of $\alpha_{\rm th}(q)$ as function of $q$ like as $\alpha_{\rm in}(q)$ in (\ref{sc-in}).
In addition, the small unstable black appears when the bound satisfies
 \begin{equation}
 r_+<r_{\rm c}(q=0.7)=1.714 \label{smubh}
 \end{equation}
 at $\alpha=8.019$.
\section{Static scalar  perturbation}

Here, it is worth  checking the instability bound (\ref{inst-b}) again because  the precise value of  $\alpha_{\rm th}(q)$ determines  scalarized RN black holes.
This can be achieved by obtaining  the static perturbed solutions to the linearized equation (\ref{sch-2}) with $\omega=0$
on the RN  background.
 For a given $l=0$ and $q$, requiring an asymptotically vanishing condition ($\varphi_\infty \to 0$) leads to the fact that  the existence  of a smooth scalar  determines  a discrete set for $\alpha$.
In addition, it determines $n=0,1,2,\cdots$ branches of scalarized black holes.
Introducing a static condition ($\omega=0$)
and a new coordinate of $z = r/2M$, the  equation  for  $u(r)$
reduces to
\begin{equation}\label{phi-r}
f(z)u''(z)+f'(z)u'(z)-\left(\frac{\alpha q^2}{4z^4}-\frac{f'(z)}{z}\right)u(z)=0,
\end{equation}
where $f(z)=(z-z_-)(z-z_+)/z^2$ with $z_{\pm}=(1\pm\sqrt{1-q^2})/2$.

Here we wish   to find a numerical solution even though  an  analytic solution is available for $l=0$ case~\cite{Herdeiro:2018wub}.
For this purpose, we first propose the near-horizon expansion for $u(z)$ as
\begin{equation}\label{expan-phi0}
u(z)=u_{+}+u'_{+}(z-z_+)+\frac{u''_{+}}{2}(z-z_+)^2+\cdots.
\end{equation}
This expression  can be used to set data  outside the outer horizon  for a numerical
integration from $z=z_+$ to $z=\infty$. Here the coefficients $u'_+$ and $u''_+$ could  be determined  in terms of
a free parameter $u_+$ as
\begin{equation}
u'_+= -\frac{\alpha q^2+4z_+(z_{-}-z_+)}{4z_+^2(z_{+}-z_-)}u_+, \quad u''_+=\frac{\alpha q^2\left(\alpha q^2+8z_-z_+\right)}{32z_+^4\left(z_+-z_-\right)^2}u_+.
\end{equation}
An asymptotic form of
$u(z)$ near the infinity of $z=\infty$ is given by
\begin{equation}\label{expan-phi1}
u(z)=u_{\infty}+\frac{u^{(1)}}{z}+\frac{u^{(2)}}{z^2}+\cdots,
\end{equation}
where  two relations are expressed in terms of $u_\infty$ as
\begin{equation}
u^{(1)}=\frac{z_{-}+z_+}{2}u_\infty,~~ u^{(2)}=\frac{-\alpha q^2+8(z_-^2+z_-z_{+}+z_+^2)}{24}u_\infty.
\end{equation}
\begin{figure*}[t!]
   \centering
   \includegraphics{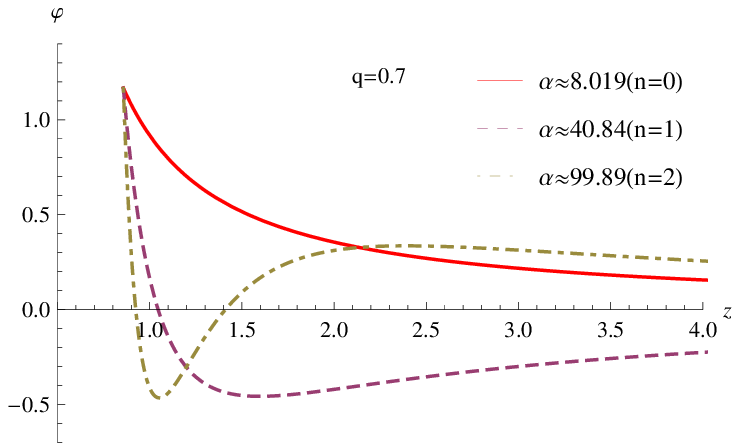}
   \hfill%
    \includegraphics{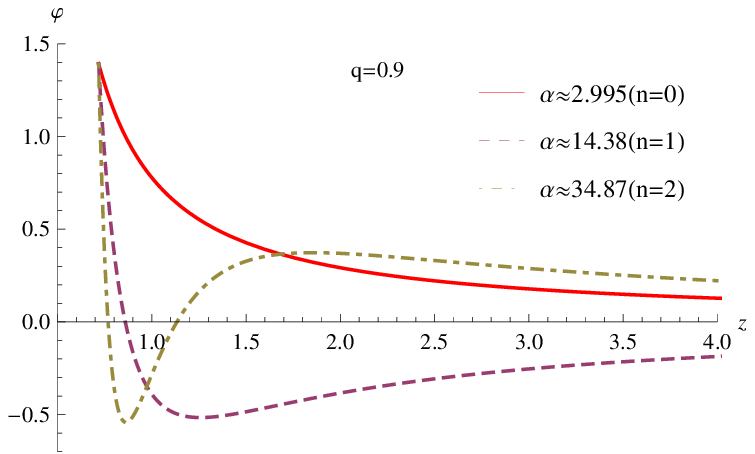}
\caption{Radial profiles of $\varphi=u(z)/z$ as function of $z=r/(2M)$ for the first three perturbed scalar solutions. The left-handed picture  is depicted for $q=0.7$, while the right-handed one is designed for $q=0.9$ (near-extremal black hole). The number of nodes $n$ is number of zero crossings.
All profiles  approach zero as $z \to \infty$.   }
\end{figure*}

 At this stage, it is worth noting  that we search for bound state scalar solution to (\ref{phi-r}) in the RN spacetime.
We are free to choose the value of scalar field $u_+$ at the horizon because (\ref{phi-r}) is a linear differential equation and then, we choose $u_\infty=1$ at infinity~\cite{Silva:2017uqg}. 
Actually, a numerical solution could be obtained  by
connecting the near-horizon form (\ref{expan-phi0}) to the asymptotic form (\ref{expan-phi1}) together with selecting the
parameter $\alpha$ for given $q$ properly. In this case, we obtain two discrete spectra of the  parameter $\alpha$:  $\alpha_n(q=0.7)\approx\{\underline{8.019}, 40.84, 99.89,\cdots\}$ and
 $\alpha_n(q=0.9)\approx \{\underline{2.995}, 14.38, 34.87,\cdots\}$. The other four spectra are given by
 $\alpha_n(0.55) \approx  \{\underline{15.46}, 80.02, 196.1,\cdots \}$,
 $\alpha_n(0.418)\approx \{\underline{29.47}, 153.9, 377.7,\cdots\}$,
 $\alpha_n(0.3) \approx \{\underline{60.69}, 318.4, 382.0,\cdots\}$,
and  $\alpha_{n}(0.25)\approx \{\underline{88.98}$, 467.4, 1148,$\cdots \}$.
In Fig.~4, these solutions are classified by the order number $n=0, 1, 2,\cdots$ which is identified with
the number of nodes for $\varphi(z) = u(z)/z$.  We find that
 the $n=0$ scalar mode without zero crossing represents the fundamental branch of scalarized black holes,
while the $n=1,2$ scalar modes  with zero crossings denote $n=1,2$ higher branches of scalarized black holes.
Actually, this corresponds to finding the $l=0$ bifurcation points from the RN black hole with $q=Q/M$.
Finally, we confirm that for given $q$,  $\alpha_{n=0}(q)=\alpha_{\rm min}(q)$ [underline value] recovers the threshold of instability $\alpha_{\rm th}(q)$  exactly.

\section{GL instability}
 The instability of the  RN black hole may be  regarded as the GL instability since this instability is based on the $s(l=0)$ mode of a perturbed scalar and its linearized equation includes an effective mass term (not tachyonic mass of $m^2_{\rm t}<0$ presicely) which develops negative potential near the horizon from the Maxwell kinetic term.
 In this section, we wish to clarify the  similarity and difference between  the GL instability (modal instability) and tachyonic instability
 because the instability of RN black hole is closely related to appearance of scalarized RN black holes.

 Let us first introduce the tachyon propagation with mass squared $m^2_t<0$ in the RN background as
 \begin{equation}\label{tach-1}
 \Big(\bar{\square}- m^2_t\Big) \varphi_t= 0
 \end{equation}
  which provides a Schr\"{o}dinger equation for radial part
  \begin{equation} \label{tach-2}
\frac{d^2u_t}{dr_*^2}+\Big[\omega^2-V_t(r)\Big]u_t(r)=0
\end{equation}
with the tachyon potential $V_t(r)$
\begin{equation} \label{tach-3}
V_t(r)=f(r)\Big[\frac{2M}{r^3}+\frac{l(l+1)}{r^2}-\frac{2Q^2}{r^4}+m^2_t\Big].
\end{equation}
As is shown Fig. 5, the potential $V_t(r)$ for $l=0$ tachyonic mode develops a positive region near horizon, while it approaches $-0.01$ as $r\to \infty$ for $m^2_t=-0.01$.
This shows clearly  the tachyonic instability of RN black hole because the sufficient condition for instability ($\int^\infty_{r_+} dr V_t(r)/f(r)=-\infty<0$)
is always satisfied with any mass $m^2_t=-{\rm const}<0$.  We wish to mention that $V_t$ differs from $V(r)$ in (\ref{pot-c}) in the sense that the latter is negative near horizon and becomes positive after crossing the $r$-axis. We regard `$-\alpha Q^2/r^4$' in $V(r)$ as an effective mass term which can be made sufficiently negative by choosing $\alpha$, making the scalar potential sufficiently negative in the near horizon. However, its role is limited to small $r$, because it approaches zero as $r\to \infty$. Such a $r$-dependent mass term is necessary to have the scalarized RN black holes.

\begin{figure*}[t!]
   \centering
  \includegraphics{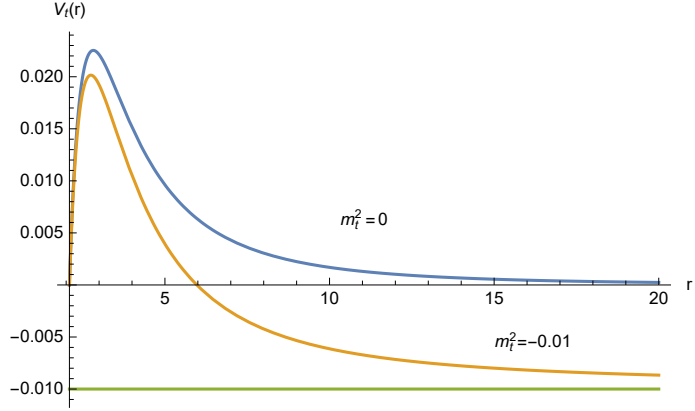}
\caption{The tachyonic  potential $V_t(r)$ as function of $r\in [r_+,\infty)$  for the outer horizon radius $r_+=2.09(q=Q/M=0.418)$ and $l=0$. For comparison, we include
a massless scalar potential with $m^2_t=0$.   }
\end{figure*}

  Now we consider the stability of Schwarzschild black hole in Einstein-Weyl gravity whose action takes the form~\cite{Myung:2013doa,Stelle:2017bdu}
  \begin{equation} \label{EW-a}
  S_{\rm EW}=\gamma \int d^4x \sqrt{-g}\Big[R-\frac{1}{2m^2_2}C_{\mu\nu\rho\sigma}C^{\mu\nu\rho\sigma}\Big].
    \end{equation}
    Its linearized equation around the Schwarzschild black hole is given by  the Licherowicz-Ricci tensor equation
    \begin{equation}\label{EW-leq}
    (\Delta_{\rm L}+m_2^2)\delta R_{\mu\nu}=0,
    \end{equation}
    where the Lichnerowicz operator is given by
    \begin{equation}
     \Delta_{\rm L}\delta R_{\mu\nu}=-\bar{\square}\delta R_{\mu\nu}-2\bar{R}_{\mu\rho\nu\sigma}\delta R^{\rho\sigma}.
     \end{equation}
     We note here that the condition of non-tachyonic mass requires $m^2_2>0$ because the Lichnerowicz operator contains $-\bar{\square}$.
     Before we proceed, we would like  to mention the GL instability. For this purpose,  we consider the perturbations around the 5D black string with $ds^2_5=ds_4^2+dz^2$ where $ds_4^2$ denotes the Schwarzschild line element,
     \[
h_{MN}(t,r,\theta,\chi,z) = \begin{bmatrix}
  h^{(4)}_{\mu\nu}& h_{\mu z}  \\
h_{z \nu} & h_{zz}
\end{bmatrix},
\]
where the $z$-dependence is assumed to be of the form $e^{ikz}$ and the time-dependence takes the form of $e^{\Omega t}$.
     Actually, Eq.(\ref{EW-leq}) takes the same form as the linearized black string equation for $h^{(4)}_{\mu\nu}$ with the transverse-traceless gauge~\cite{Gregory:1993vy}
     \begin{equation}
     (\Delta_{\rm L}+k^2)h^{(4)}_{\mu\nu}=0
     \end{equation}
     except that  the mass $m_2$ of the Ricci tensor is replaced by the wave number $k$ along $z$ direction.
     The GL instability states that the 5D black string is unstable against the metric perturbation  for $k<k_{\rm th}=0.876/r_+$ (long wavelength perturbation).
  The GL instability is an $s(l=0)$-wave spherically symmetric instability from the four-dimensional perspective.
      In addition, it is interesting to note that the dRGT massive gravity having a Schwarzschild solution  when formulated in a diagonal bimetric form, has the same linearized equation as  (\ref{EW-leq}) except replacing $\delta R_{\mu\nu}$ by $h_{\mu\nu}$~\cite{Babichev:2013una,Brito:2013wya}.

     The $l=0$ polar sector of Eq.(\ref{EW-leq}) is given by
      \begin{equation} \label{EM-s}
\frac{d^2\tilde{\varphi}_0}{dr_*^2}-\Big[\Omega^2+V_Z(r)\Big]\tilde{\varphi}_0(r)=0
\end{equation}
with $\tilde{\varphi}_0$=$s(l=0)$-mode of $\delta R_{\mu\nu}$ and the Zerilli potential $V_Z(r)$~\cite{Brito:2013wya,Lu:2017kzi}
\begin{equation} \label{EW-p}
V_Z(r)=\Big(1-\frac{r_+}{r}\Big)\Bigg[\frac{r_+}{r^3}+m^2_2-\frac{12r_+(r-0.5r_+)m^2_2+6r^3(2r_+-r)m^4_2}{(r_++r^3m^2_2)^2}\Bigg].
\end{equation}
As is shown in Fig. 6, all potentials develop negative region near the horizon, whereas their asymptotic limits are nonzero constants ($V_Z \to m^2_2$, $r\to \infty$).
The former is similar to $V(r)$ in (\ref{pot-c}), while the latter is different from $V\to 0$ as $r\to\infty$.  This may imply that the structure of scalarized black holes
differs from that of non-Schwarzschild black hole (Schwarzschild black hole with Ricci-tensor hair).
\begin{figure*}[t!]
   \centering
  \includegraphics{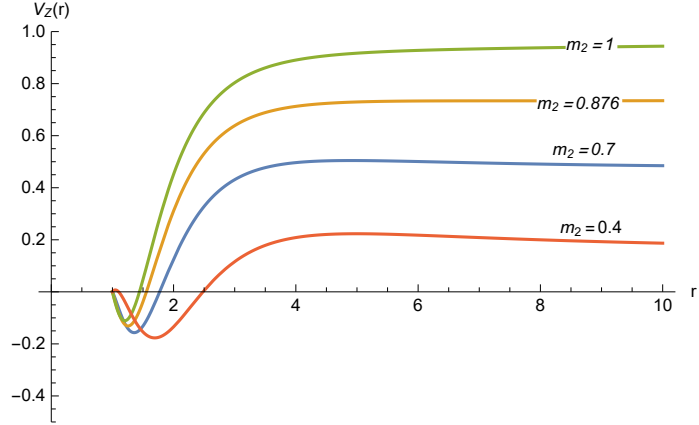}
\caption{The Zerilli  potentials $V_Z(r)$ as function of $r\in [r_+,\infty)$ with different masses $m_2$=1(stable) ,0.876(threshold), 0.7(unstable), 0.4(unstable)  for the horizon radius $r_+=2M=1$ and $l=0$.   }
\end{figure*}
Solving Eq.(\ref{EM-s}) with boundary conditions, one finds unstable tensor modes from Fig. 7.
\begin{figure*}[t!]
   \centering
   \includegraphics{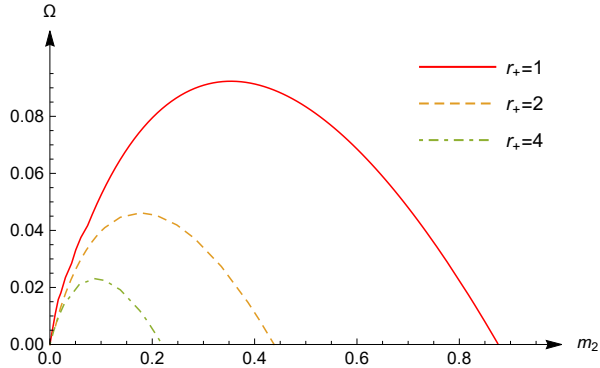}
\caption{Plots of unstable tensor modes with different horizon radii.
The $y(x)$-axis denote $\Omega$ in $e^{\Omega t}$ (mass $m_2$ of massive spin-2 mode).  Here we read off the thresholds of instability  located at $m^{\rm th}\approx$0.876,~0438,~0.219 for $r_+=1,2,4$.}
\end{figure*}
From Fig. 7, the GL instability mass bound for $s(l=0)$-mode is given by
\begin{equation} \label{GL-bound}
0<m_2<m_{\rm th}=\frac{0.876}{r_+},
\end{equation}
where $m_{\rm th}$ represents the threshold of GL instability.

On the other hand, we confirm the precise value of $m_{\rm th}$ by solving the static Lichnerowicz-Ricci tensor equation as~\cite{Lu:2017kzi}
\begin{equation} \label{SLR-eq}
\Delta_{\rm L}\psi_{\mu\nu}=\lambda \psi_{\mu\nu},
\end{equation}
where the eigenvalue $\lambda$ should be determined by requiring the existence of a normalizable eigenfunction $\psi_{\mu\nu}$.
This amounts to seeking a negative  eigenvalue $\lambda$ for which the exponentially diverging solution $e^{\sqrt{-\lambda}}$ is absent when solving (\ref{SLR-eq}) [equivalently, (\ref{EM-s}) with $\omega=0$ and $m^2_2=-\lambda$]
numerically by the shooting method.
It determines   $\lambda=-m^2_2=-0.7677$, leading to $m_{2}=m_{\rm th}=0.876$.
We emphasize that there exists just one negative value of $\lambda$ for which one can have a normalizable eigenfunction.
Here, we note that  $\lambda=-0.7677$ is relevant both to the edge of the zone of Schwarzschild instability and  the existence of non-Schwarzschild black holes.
Importantly, this process is very similar to  Section 4  for determining $\alpha_{\rm th}$ in the EMS theory.
The difference is that  many branches of $\alpha=\{8.019,40.84,99.89,\cdots\}$ for $q=0.7$ exist in the EMS theory, while a single branch of $m^2_2=0.7677$ exists for the EW gravity.
This may be so because their asymptotic forms of potentials are different ($V\to 0$ versus $V_Z \to m^2_2$ as $r\to \infty$).
Hence, the boundary condition at infinity  is an asymptotically vanishing scalar ($\varphi_\infty \to 0$) in the EMS theory, while it is a normalizable mode in the EM gravity.
An actual correspondence would be met if one includes a mass term of $V(\phi)= 2\alpha \phi^2$ in (\ref{Action1}), leading to the $s(l=0)$-mode potential
\begin{equation} \label{cpot-c}
V_{\rm cp}(r)=f(r)\Big[\frac{2M}{r^3}+\alpha-(\alpha+2)\frac{Q^2}{r^4}\Big],
\end{equation}
which has  similar asymptote ($V_{\rm cp} \to \alpha$ as $r\to \infty$)  to $V_Z(r)$ in (\ref{EW-p}).

From (\ref{GL-bound}),  selecting $m_{\rm th}=1$ for $r_+=r_c=0.876$, one finds the bound for unstable (small) black holes
\begin{equation}
r_+<r_c.
\end{equation}
It is worth noting that  $r_+=r_c$ corresponds to the bifurcation point which allow   a new non-Schwarzschild black hole~\cite{Lu:2015cqa}.
At this stage, we note that the appearance of non-Schwarzschild black hole is closely related to the threshold of instability for Schwarzschild black hole in the Einstein-Weyl gravity~\cite{Myung:2013doa,Stelle:2017bdu}.

We summarize whole properties for  instability happened in the  EMS theory and Einstein-Weyl gravity in Table 1.
It is emphasized  that  the role of $s$-mode scalars $\varphi$ in the EMS theory
is replaced by a $s$-mode Ricci tensor $\delta R_{\mu\nu}(\tilde{\varphi}_0)$ in the Einstein-Weyl gravity.

\begin{table}[h]
\resizebox{\textwidth}{20mm}
{\begin{tabular}{|c|c|c|}
  \hline
   Theory &Einstein-Maxwell-scalar theory &  Einstein-Weyl gravity\\ \hline
  Action & $S_{\rm EMS}$ in (\ref{Action1})  & $S_{\rm EW}$ in (\ref{EW-a})   \\ \hline
  BH without hair & RNBH with $\bar{\phi}=0$ &SBH with $\bar{R}_{\mu\nu}=0$ \\ \hline
  Linearized equation & scalar equation (\ref{l-eq3})  & LR-equation (\ref{EW-leq}) \\ \hline
  GL instability mode & $s$-mode of $\varphi$ &$s$-mode of $\delta R_{\mu\nu}$ \\ \hline
  Bifurcation points &$\alpha=8.019,40.84,99.89,\cdots$ for $q=0.7$& $m_2^2=0.7677$ \\ \hline
  Potential and its asymptotic form &$V(r)$ in (\ref{pot-c}) and $V_{r\to\infty}=0$   &$V_Z(r)$ in (\ref{EW-p}) and  $V_{Z,r\to\infty}=m_2^2$ \\ \hline
  GL instability bound &$\alpha>\alpha_{\rm th}(q)$ in (\ref{inst-b})  & $0<m_2<\frac{0.876}{r_+}$ in (\ref{GL-bound}) \\ \hline
  Small unstable BH & $r_+<r_c=1.714$ with $\alpha_{\rm th}=8.019(q=0.7)$  &$r_+<r_c=0.876$ with $m_{\rm th}=1$ \\ \hline
  BH with hair & scalarized RN  BH    & non-Schwarzschild BH \\ \hline
\end{tabular}}
\caption{Gregory-Laflamme (GL) instability among  RN black hole (RNBH) in EMS theory and
 and Schwarzschild black hole (SBH) in Einstein-Weyl gravity. LR denotes  Licherowicz-Ricci tensor. }
\end{table}

\section{Scalarized RN black holes}
\subsection{Exponential coupling}
Before we proceed, we note  that the RN black hole solution  is allowed for any value of $\alpha$,
while a scalarized RN black hole solution
may exist only for $\alpha\ge \alpha_{\rm th}$.
The threshold of  instability for a RN black hole  reflects the disappearance of zero crossings in the perturbed scalar profiles.
We explore a close connection between the instability of a RN black hole without scalar hair and appearance of  a scalarized RN black hole.
As a concrete  example, we wish  to find  a scalarized RN black hole which is closely related to the  $q=0.7(M=1,~Q=0.7)$ and $\alpha \ge 8.019$ case ($n=0$ case).

For this purpose, let us introduce  the metric ansatz as~\cite{Herdeiro:2018wub}
\begin{eqnarray}\label{nansatz}
ds^2_{\rm sRN}=-N(r)e^{-2\delta(r)}dt^2+\frac{dr^2}{N(r)}+r^2(d\theta^2+\sin^2\theta d\chi^2),
\end{eqnarray}
where a metric function is defined by $N(r)=1-2m(r)/r$ with the mass function $m(r)$. Also, we consider the $U(1)$ potential and the scalar   as $A=v(r)dt$ and $\phi(r)$.
Substituting these  into Eqs.(\ref{equa1})-(\ref{s-equa}) leads to  the four equations
\begin{eqnarray}\label{neom}
&&-2m'(r)+e^{2\delta(r)+\alpha\phi(r)^2}r^2v'(r)^2+[r^2-2rm(r)]\phi'(r)^2=0,\label{neom1}\\
&&\delta'(r)+r\phi'(r)^2=0,\label{neom2}\\
&&v'(r)+e^{-\delta(r)-\alpha\phi(r)^2}\frac{Q}{r^2}=0,\label{neom3}\\
&&e^{2\delta(r)+\alpha\phi(r)^2}r^2\alpha\phi(r)v'(r)^2+r[r-2m(r)]\phi''(r)\nonumber\\
&&-\Big(m(r)[2-2r\delta'(r)]
+r[-2+r+2m'(r)]\delta'(r)\Big)\phi'(r)=0. \label{neom4}
\end{eqnarray}

Assuming the existence of a horizon located at $r=r_+$,  one finds an
approximate solution to equations in the near horizon
\begin{eqnarray}\label{nexpr}
&&m(r)=\frac{r_+}{2}+m_1(r-r_+)+\ldots,\label{aps-1}\\
&&\delta(r)=\delta_0+\delta_1(r-r_+)+\ldots,\label{aps-2}\\
&&\phi(r)=\phi_0+\phi_1(r-r_+)+\ldots,\label{aps-3}\\
&&v(r)=v_1(r-r_+)+\ldots,\label{aps-4}
\end{eqnarray}
where the four coefficients are given by
\begin{eqnarray}\label{ncoef}
 m_1=\frac{e^{-\alpha\phi_0^2}Q^2}{2r_+^2},\quad
\delta_1=-r_+\phi_1^2,\quad \phi_1=\frac{\alpha\phi_0 Q^2}{r_+(Q^2-e^{\alpha\phi_0^2}r_+^2)},\quad v_1=-\frac{e^{-\delta_0-\alpha\phi_0^2}Q}{r_+^2}.
\end{eqnarray}
This approximate solution involves  two  parameters of  $\phi_0=\phi(r_+)$ and $\delta_0=\delta(r_+)$, which will be
found when  matching  (\ref{aps-1})-(\ref{aps-4}) with the asymptotic  solutions in the far region
\begin{eqnarray}\label{ncoef}
m(r)&=&M-\frac{Q^2+Q_s^2}{2r}+\ldots,~\phi(r)=\phi_\infty+\frac{Q_s}{r}+\ldots, \nonumber \\
\delta(r)&=&\frac{Q_s^2}{2r^2}+\ldots,~v(r)=\Phi+\frac{Q}{r}+\ldots, \label{insol}
\end{eqnarray}
where  $Q_s$ and $\Phi$ denote the scalar charge and the electrostatic potential, in addition to the ADM mass $M$ and the electric charge $Q$.
For simplicity, we choose  $\phi_\infty=0$ here.

The EMS theory admits the RN black hole solution  for any $\alpha$. However, it becomes an unstable black hole for $\alpha>\alpha_{\rm th}(q)$ (\ref{inst-b}), while it is stable against the scalar perturbation for $\alpha<\alpha_{\rm th}(q)$. We note that  `$\alpha=\alpha_{\rm th}(q)$' indicates the threshold of instability.
One expects  that a scalarized RN black hole is allowed  for $\alpha \ge\alpha_{\rm th}(q)$ when $q\ge0.7$.  This means that the scalarized RN black holes bifurcates from the RN black hole hole at $\alpha =\alpha_{\rm th}(q)$, but  $q$ increases beyond unity for the fixed $\alpha$, implying that the scalarized RN black hole could be overcharged~\cite{Herdeiro:2018wub}.
\begin{figure*}[t!]
   \centering
   \includegraphics{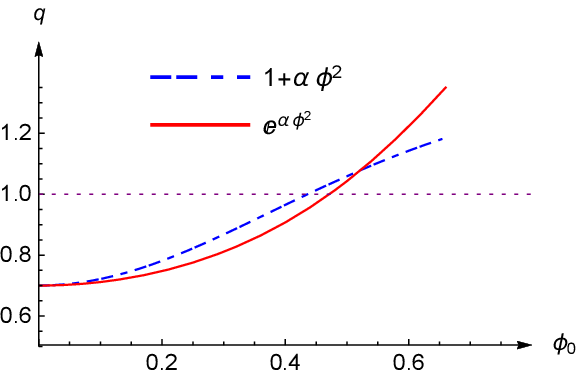}
      \hfill%
    \includegraphics{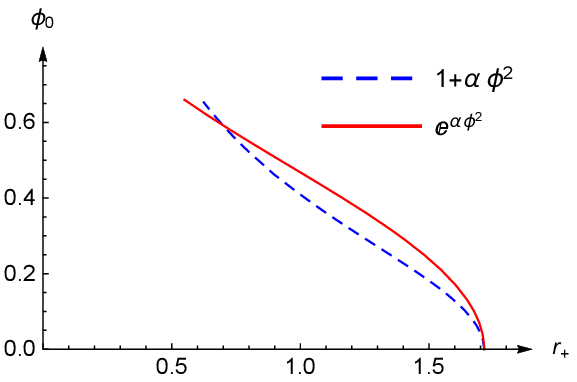}
\caption{(left) The charge-mass ratios  $q$ of the $n=0$ scalarized RN  black hole as
functions of $\phi_0$ with the fixed $\alpha=8.019$ for exponential and quadratic couplings. The  horizontal line represents the maximum ratio $q=1$ for the RN black hole.
 (right) The scalars at the horizon $\phi_0$ as  functions of the horizon radius $r_+$  for the $n=0$  scalarized black hole. }
\end{figure*}

For the RN black hole with  $\phi_0=0$, the outer horizon is located at $r_{+}=1.714$ and the charge-mass ratio is given by $q=0.7$.
In the Fig.8 (left), one observes that for given $\alpha=8.019$, the ratio of $q$ for the $n=0$  scalarized RN black hole increases  beyond the extremal  RN black hole ($q=1$) as  $\phi_0$ increases.
Moreover, in the  Fig.~8 (right), the scalar at the horizon $\phi_0$ increases as the horizon radius $r_+$  decreases. The scalar at the horizon
is terminated at $r_+=1.714$, corresponding  to the RN outer horizon. It is the starting point for  a scalarized RN black hole, while from (\ref{smubh}) it corresponds to  the ending point for unstable  RN black hole.

It is known that the scalarization bands exist for the ESGB theory~\cite{Silva:2017uqg}. A discrete set for $\eta/M^2$ obtained from static scalar perturbation  corresponds
to the right-end values of scalarization bands for a scalarized  Schwarzschild black hole, while the left-end values are provided by the regularity constraint at the horizon ($r_+^4\ge6\eta^2\phi_0^2 $). However, as is shown in Fig. 9, there are  no scalarization bands in the EMS theory
because we do not need to impose  the regularity condition at the horizon. As a result, there is no upper bound on $\alpha$  as  $n=0(\alpha\ge 8.019),1(\alpha\ge40.84),$ and $2(\alpha\ge 99.89)$.

\begin{figure*}[t!]
   \centering
   \includegraphics{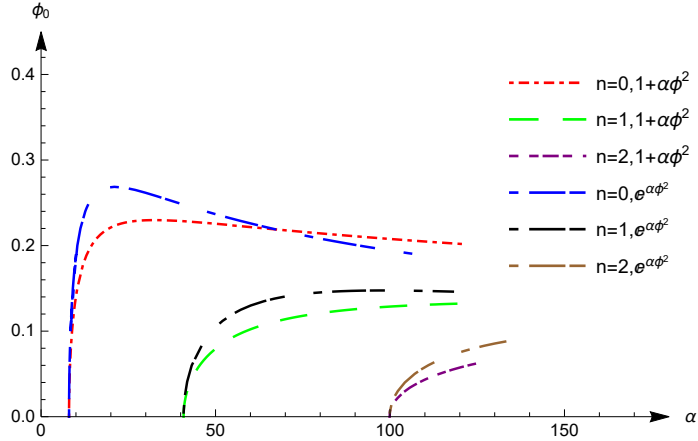}
\caption{The scalar field  $\phi_0=\phi(r_+)$ at the horizon as a function of mass-like parameter $\alpha$ for exponential and quadratic couplings . All the nontrivial branches with $n=0,1,2$ start from  bifurcation points at $\alpha_n=\{8.019, 40.84, 99.89\}$ on the trivial branch (RN black holes on $\alpha$-axis).  They span the whole region without upper bound.}
\end{figure*}

Consequently, we obtain the scalarized RN black hole solution depicted in Fig. 10.
The metric function $N(r)$ has a different horizon at $\ln r=\ln r_+=0.067$ in compared to the RN horizon at $\ln r=\ln r_+=0.539$ and
it approaches the RN metric function $f(r)$ as $\ln r$ increases. Also,  the scalar hair $\phi(r)$ starts with  $\phi_0=0.44$ at the horizon and it decreases as $\ln r$ increases, in compared to $\phi(r)=0$ for the RN black hole.

\begin{figure*}[t!]
   \centering
   \includegraphics{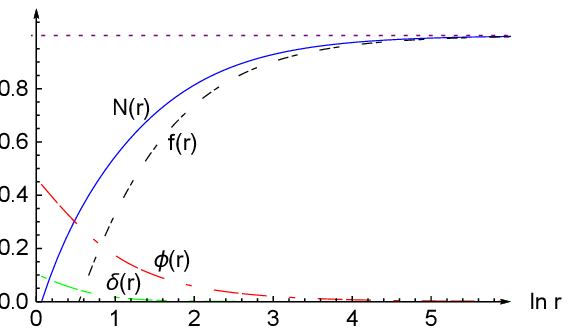}
      \hfill%
    \includegraphics{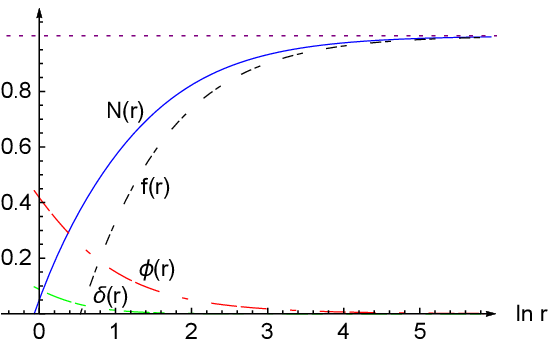}
\caption{The $n=0$  scalarized RN solutions. (left) Exponential coupling.  Two metric functions $f(r)$ for RN  and  $N(r)$ for scalarized RN are given by  functions of   $r$  for $\alpha=8.019$.
We observe that the logarithmic values of horizon radius of  scalarized  RN and RN  black holes are
located at 0.067 and  0.539, respectively.
 The scalar hair $\phi(r)$ starts with  $\phi_0=0.44$ at the horizon and it decreases as $\ln r$ increases.  (right) Quadratic coupling.
 The logarithmic values of horizon radius of  scalarized  RN and RN  black holes are
located at $-0.062$ and  0.539, respectively.
 }
\end{figure*}

\subsection{Quadratic coupling}
Considering  the quadratic coupling of $\alpha \phi^2$, we have to choose $\bar{\phi}={\rm const}$ to obtain the RN black hole with different charge $\tilde{Q}^2=\alpha \bar{\phi}^2 Q^2$.
In order to make the analysis simple, we may  choose an equivalent coupling of  $1+\alpha \phi^2$ with $\bar{\phi}=0$ to give the RN black hole.
In this case, the bifurcation points of the RN solution are the same as those of exponential coupling because the static scalar equation takes the same form as in
(\ref{phi-r}). Furthermore, instabilities of RN solution are exactly the same for both couplings. To obtain  scalarized RN black holes, we solve Eqs.(\ref{neom1})-(\ref{neom4})
by replacing  $e^{\alpha \phi^2}$ with  $1+\alpha \phi^2$. From Figs. 8, 9, 10, we observe that the quadratic coupling shows the similar properties to the exponential coupling.

As was mention in~\cite{Blazquez-Salcedo:2018jnn}, however, the only difference between two coupling in the ESGB theory is that the $n=0$ fundamental branch  of scalarized black holes is stable for the exponential coupling, while the $n=0$ fundamental branch is unstable for the quadratic coupling.
Therefore, we expect that the similar thing will happen since the $n=0$ scalarized RN  black hole turned out to be unstable in  the EMS theory with exponential coupling~\cite{Myung:2018jvi}.

\section{Discussions}

First of all, we mention that  scalarized RN black holes were found in the EMS theory.
It is emphasized that the appearance of these black holes with scalar hair is closely related to the instability of the RN black hole without scalar hair in the EMS theory.
Concerning the appearance scalarized RN black holes~\cite{Herdeiro:2018wub}, it is very important to obtain the precise threshold $\alpha_{\rm th}$ of instability for the RN black hole in the  EMS theory. In this work, we have obtained the GL instability bound (\ref{inst-b}) for the RN black hole in the EMS theory by considering $s(l=0)$-mode scalar perturbation.

Roughly speaking,  a shape of  scalar potential $V(r)$ in (\ref{pot-c})  determines the  instability of RN  black hole.  The sufficient condition of $\int^{\infty}_{r_+}
 dr [V(r)/f(r)]<0$ for instability~\cite{BS,Dotti:2004sh}
gives rises to an analytic bound (\ref{sc-in}), while the sufficient condition for stability is given by the other bound (\ref{c-po}). Explicitly, for $q=0.418$,  the sufficient condition for instability takes the form of $\alpha>30.74$, whereas the sufficient  condition for the stability is given by $0<\alpha \le 19.83$.
In the case of $\int^{\infty}_{r_+}dr [V(r)/f(r)]>0$ with negative potential near the horizon, however, it is not easy to make a clear decision on the stability of the black hole. Here it is still stable  for $19.83<\alpha\le29.47$ with $q=0.418$ even providing negative region near the horizon shown in Fig. 1.  In this case,  the $S$-deformation method might provide a complementary result to  support  the stability of such black holes by finding the deformed potential~\cite{Kimura:2017uor,Kimura:2018eiv}.

In general, the GL instability bound  is not  given by an analytic form.  As was shown in Fig. 3 depending on $q$, it was determined by solving the linearized  equation (\ref{l-eq3})  numerically. In the case of $q=0.418$, the GL instability bound is $\alpha > 29.47$  which is surely less than the sufficient condition for instability
($\alpha>30.74$).  Importantly, this picture shows that  the GL instability  appeared  in a simpler EMS theory than the ESGB theory and Einstein-Weyl gravity.
For $q=0.7$, we have obtained the GL instability bound of $\alpha >\alpha_{\rm th}= 8.019$.
We have  derived  the precise value of threshold  $\alpha_{\rm th}=8.019$  again by solving  the static linearized  equation numerically.
Furthermore, we have obtained the $n=0(\alpha\ge8.019)$ scalarized RN black hole  by solving  Eqs.(\ref{neom1})-(\ref{neom4}) numerically for exponential and quadratic couplings.

Consequently, we have explored   a clear connection between GL instability of RN black hole and  scalarization of RN black hole.

 \vspace{1cm}

{\bf Acknowledgments}
 \vspace{1cm}

This work was supported by the National Research Foundation of Korea (NRF) grant funded by the Korea government (MOE)
 (No. NRF-2017R1A2B4002057).

\end{document}